\documentclass[aps,prb,floatfix,superscriptaddress,twocolumn,footinbib]{revtex4-1}
\usepackage{graphicx}
\usepackage{dcolumn}
\usepackage{bm}
\usepackage{makecell}
\usepackage{color}
\usepackage{amsmath}
\usepackage{latexsym}
\usepackage{amssymb}
\usepackage{graphics,epstopdf}
\usepackage{hyperref}
\usepackage{xcolor}
\hypersetup{
    colorlinks = true,
    linkcolor =blue,
	citecolor=blue,
	urlcolor=blue
}

\newcommand{\ed}{\end{document}}
\newcommand{\beq}{\begin{equation}}
\newcommand{\eeq}{\end{equation}}

\usepackage{mathtools}
\usepackage{comment}

\begin{document}

\title{Spin-dependent destructive quantum interference associated with chirality-induced spin selectivity in circular single helix molecules}

\author{Song Chen}
 \affiliation{School of Physics and Wuhan National High Magnetic Field Center,
Huazhong University of Science and Technology, Wuhan 430074, People's Republic of China.}

\author{Hua-Hua Fu}
\altaffiliation{Corresponding author.\\ hhfu@hust.edu.cn}
\affiliation{School of Physics and Wuhan National High Magnetic Field Center,
Huazhong University of Science and Technology, Wuhan 430074, People's Republic of China.}
\affiliation{Institute for Quantum Science and Engineering,
 Huazhong University of Science and Technology, Wuhan, Hubei 430074, China.}

\date{\today}

\begin{abstract}
Theoretical studies on spin-dependent transport through helical molecules with straight spiral geometry have received intense research interest in the past decade, however, the physics in circular helical molecules has still less been explored. In this work, we theoretically construct a circular single helix (CSH) possessing the chirality-induced spin-orbit coupling and contacting with two non-magnetic electrodes. Our theoretical calculations demonstrate that the spin-related transport in CSH exhibits the so-called chiral-induced spin selectivity (CISS) effect and more importantly, the CISS-reduced spin-dependent destructive quantum interference (DQI) also occurs in the CSH, without any external magnetic field or magnetic electrodes. Moreover, the changing of CSH length or the electrode positions exhibits specific patterns in the spin-polarized conductance. Particularly, the dephasing magnitude can adjust effectively these two spin-dependent effects to realize their coexistence. Additionally, the phase transition between the CISS-dependent constructive quantum interference (CQI) and DQI is also observed in the CSH. Our theoretical work puts forwards a new material plateau to explore the CISS and to exhibit the novel CISS-dependent CQI effect.

\end{abstract}

\maketitle

\section{INTRODUCTION}

Chirality-induced spin selectivity (CISS) is a fascinating phenomenon which has attracted much attention in recent decades due to its potential applications in spintronics and molecular electronics. This effect refers to the preferential transmission of electrons with a particular spin orientation through chiral molecules, which exhibit a non-superimposable mirror image. To understand the physical mechanism of CISS observed in the helix DNA molecules, many model Hamiltonians based on the straight helical geometry have been proposed by involving the spin-orbit coupling (SOC) in theory \cite{GuoSun2012, EremkoLoktev2013, GuoSunPRB2014, GuoSun2014, Matityahu2016,Dalum2019, Daz2018, Fransson2019, Geyer2019,Michaeli2019, DuFu2020,Yan2021}.

The configuration of biomolecules in a helical geometry may not always be accurately depicted as a simple straight helix. As an instance, DNA frequently undergoes curvatures when it is part of a molecular compound, such as with proteins and in chromatin. A viral DNA in a circular arrangement has also been observed in nature. Moreover, cyclic proteins, which have been found in various organisms including bacteria, plants, and animals, are a type of protein in which the amino (N) and carboxy (C) termini are chemically connected together to form a circular backbone \cite{Martinez-Bueno, Alain Blond, David J Craik}. With the development of protein nanotechnology, an increasing number of complex tandem repeat proteins, made up of repeating units of amino acid sequences and structures, can be synthesized artificially \cite{Rmisch, Boersma,Voet,Po-Ssu Huang, Derrick}. For example, Doyle $et$ $al.$ applied a completely novel approach for the $de$ $novo$ design of tandem repeat protein architectures based solely on geometric criteria, without referring to the sequences and structures of existing repeat protein families, resulting in a particular structures that is much different from those observed in nature, such as the closed $\alpha-$toroids repeat proteins with left-handed helix \cite{Lindsey Doyle} and Hicks $et$ $al.$ outlines a universal strategy for creating $C_2$ symmetric proteins which are hyperstable and contain pockets with various sizes and geometries \cite{Derrick}.

On the other side, the electronic transport in single-molecule devices may exhibits a special quantum phenomenon, i.e., the quantum interference (QI) due to multiple electron pathways interfere with each other either constructively or destructively in energy space \cite{Solomon2008,Liu2018,SaraivaSouza2014}. In particular, the destructive quantum interference (DQI) is marked by a sharp dip in the transmission function of electrons transporting through in the single-molecule devices, which may result in a significant decrease in transmission by several orders of magnitude. In contrast, the constructive quantum interference (CQI) enhances the CISS-induced spin-polarized transmission with a theoretical upper limit of 4-fold conductance. Moreover, molecular devices may exhibit two inspiring conductive modes, depending on the presence or absence of DQI, making it a vital area of research towards potential CISS-based device applications in logic gates, thermoelectric devices, molecular switches, and spin filters. Furthermore, by introducing non-magnetic electrodes, spin-dependent DQI effects can be achieved, allowing for the realization and modulation of spin-dependent transport phenomena in molecular junctions \cite{LiD2019,ShuaiQiu2021}. Interestingly, both the CISS and DQI effects existing in molecules can be observed at room temperature \cite{Waldeck2021,Ruijiao Miao} without the extremely harsh conditions required for other phenomena in condensed matter physics, such as high pressures and low temperatures.

\begin{figure}
\includegraphics[width=\columnwidth]{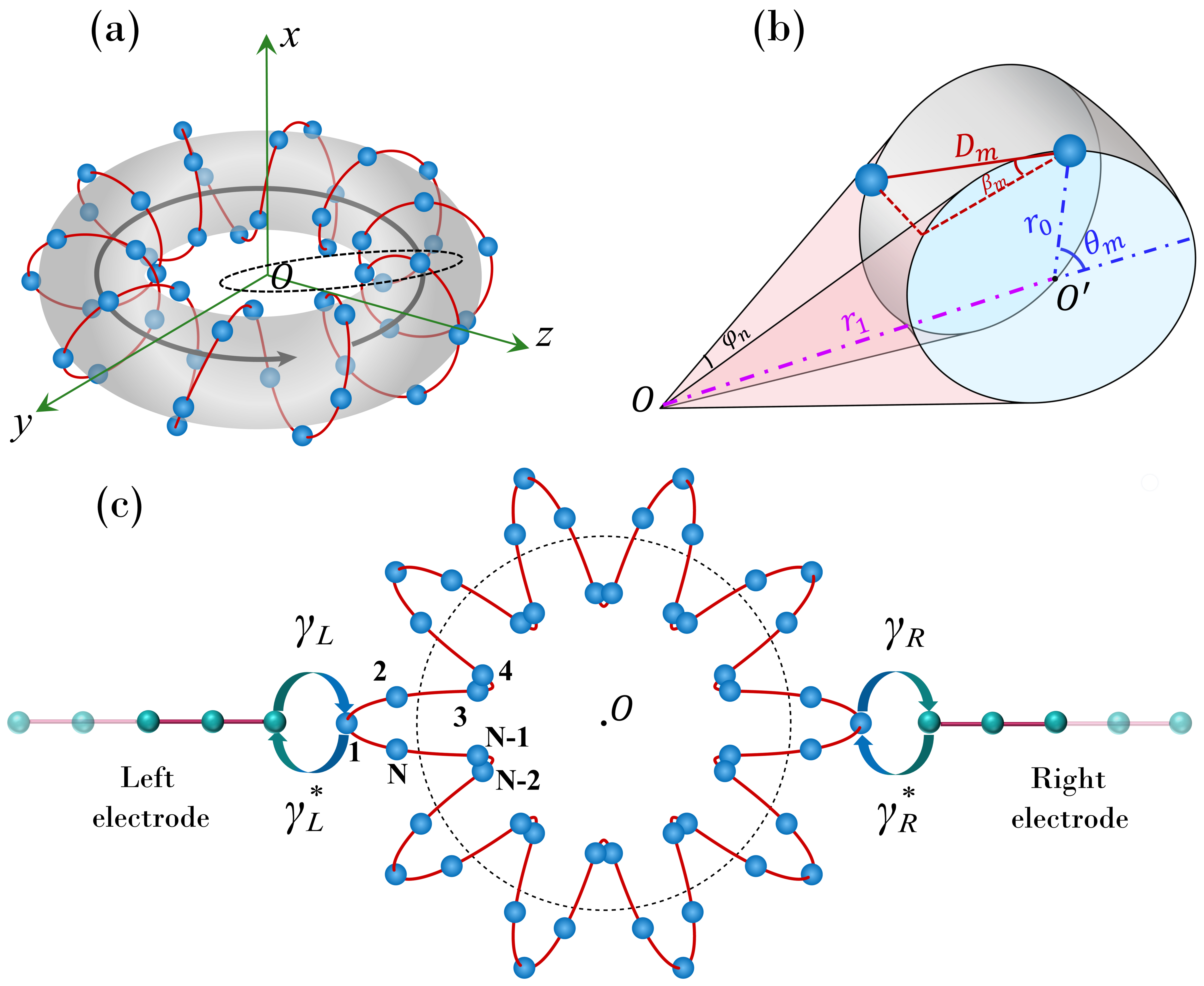}
\caption{(a) The side-view schematic illustration of the circular single helical molecule. (b) A magnified schematic diagram of the portion enclosed by the black dashed circle in (a). Here, $O$ and $r_1$ denote the center and radius of the toridal chiral molecule , $O^{'}$ and $r_0$  denote the center and radius of its cross-sectional plane. $\phi_n$ and $\theta_m$ respectively characterize the toroidal angle and poloidal angle, respectively. $\beta_m$ and $D_m$ represent space angle and the distances  between two atoms located in the cross-sectional planes of circular single helix (CSH), respectively. (c) A schematic top-view of a CSH molecule connected with two non-magnetic electrodes with coupling strengths of $\gamma_L$ ($\gamma_L^{*}$) and $\gamma_R$ ($\gamma_R^{*}$). For the leftmost atom in the central molecular system, we label it as 1, and label the remaining atoms clockwise until $\mathcal{N}$, where $\mathcal{N}$ is the length of CSH molecule.}
\label{fig1}
\end{figure}

In this paper, we propose a model Hamiltonian within a tight-binding description, including the chirality-induced SOC and the dephasing induced by environment, to study the CISS and DQI effect through a circular left-handed $\gamma$-helix protein-like molecules, which have a continuum of small curvatures and $C_2$ symmetry and are coupled with two non-magnetic electrodes, as described in Fig.~\ref{fig1}(a). By using the Landauer-B$\ddot{u}$ttiker probe (LBP) technique \cite{Kilgour2015,Kilgour2015JPCC,Kilgour2016JCP}, we calculate the spin-dependent transport properties and uncover that only when the central CSH molecule is localized in a connection state, possesses chirality and experiences environment-induced decoherence, the spin-polarized effect will be exhibited and meanwhile, a nearly perfect CISS effect can be observed. When $\mathcal{N}$ satisfies the relation $\mathcal{N} \bmod 2=0 \wedge \mathcal{N}\bmod M=0$, the spin-dependent conductance at the Fermi level exhibits DQI or CQI depending on whether the left and right electrodes are connected with the same sublattice. The number of spin-polarized zero points gradually increases from 1 to an odd number, depending on the number of lattice sites between two electrodes. However, when $\mathcal{N}$ satisfies the relation $\mathcal{N} \bmod 2=1 \wedge \mathcal{N}\bmod M=0$, no any DQI effect occurs at the Fermi level regardless of how both electrodes are connected and meanwhile, the number of spin-polarized zero points gradually increases from 0 to an even number, tightly depending on the position numbers connected by both electrodes. In addition, as the dephasing parameter $\gamma_d$ gradually increases from 0 to a finite one, the DQI effect weakens while the CISS effect is enhanced largely firstly and then weakened. Through adjusting the strengthen of SOC, the phase transition between CISS-dependent DQI and CQI occurs clearly occurs at the connection point of different odd and even electrodes only when $\mathcal{N} \bmod 4=0 \wedge \mathcal{N}\bmod M=0$.

\section{MODEL AND METHODS}

\subsection{Model Hamiltonian}

We consider a CSH molecule connected with two non-magnetic leads at its two ends, as depicted diagrammatically in Fig.~\ref{fig1}(c). The charge transport behaviors in the CSH molecule can be simulated by the following model Hamiltonian:
\begin{eqnarray}
\mathcal{H}=\mathcal{H}_{c}+\mathcal{H}_{\mathrm{so}}+\mathcal{H}_{l}
+\mathcal{H}_{lc} +\mathcal{H}_{\mathrm{p}}+\mathcal{H}_{\mathrm{pc}}.
\end{eqnarray}
The first term $\mathcal{H}_{c}$ ($=\hat{\mathbf{p}}^2/2m_e+V$) provides a description of the kinetic and potential energies associated with the conductive electrons in the CSH with $\hat{\mathbf{p}}$ the momentum operator and $m_e$ the electron mass. In the second quantization formalism, $\mathcal{H}_{c}$ is discretized as:
\begin{eqnarray}
\mathcal{H} _c=\sum_{n=1}^{\mathcal{N}}{\varepsilon _nc_{n}^{\dagger}c_n}+\sum_{n=1}^{\mathcal{N} -1}{\left(t_{n}c_{n}^{\dagger}c_{n+1}+\mathrm{H}.\mathrm{c}. \right)},
\end{eqnarray}
where, $c_{n}^{\dagger}=\left(c_{n \uparrow}^{\dagger}, c_{n \downarrow}^{\dagger}\right)$ and $\varepsilon_n$ are the creation operator and on-site energy for the electrons at the $n^{th}$ site position in the $\mathcal{N}$-lattice CSH molecule. For simplicity, $t_n$ is uniform and defined as $t$, and only the nearest-neighbor (NN) hopping is considered here.

The second term, $\mathcal{H}_{\mathrm{so}}=\frac{\hbar}{4 m_e^2 c^2} \nabla V \cdot(\hat{\sigma} \times \hat{\vec{p}})$, describes the SOC, which is an relativistic interaction between an electron with a spin index and its motion in a potential field. Here, $V$ represents electrostatic potential, $\hbar$ is the reduced Planck constant, $c$ is the speed of light, $\hat{\sigma}$ is Pauli vector and $\vec{p}$ is the three-dimensional momentum operator.
Firstly, since the variation of $V$ is almost rapid in the vicinity of the nucleus, it is reasonable to only consider its radial component $\hat{r}$. Secondly, we consider at the boundary where the radius $r_0$ is located, the value of $\nabla V$ is significantly large \cite{GuoSun2012}. Finally,  only the torus structure surrounding the CSH molecule is approximate to consist of some very small straight cylindrical units for determining the relations of various positional parameters. And for a special CSH molecule, such as a circular DNA, as the number of base pairs increases, the DNA's curvature rapidly decreases, even if the total number of base pairs is large to 100 \cite{Bhattacherjee}. Moreover, under the conditions of low curvature and long CSH chain, we believe that the numerical error may be minimized as much as possible.
Now let us introduce the covariant basis vectors to concretize $\mathcal{H}_{\mathrm{so}}$ in toroidal coordinates as $\mathcal{H}_{\mathrm{so}} = \frac{\partial}{\hbar}\left[ -i\hbar \frac{1}{\left( r_1+r_0\cos \theta \right)}\sigma _{\theta}\frac{\partial}{\partial \varphi}-\frac{1}{r_0}\sigma _{\varphi}\frac{\partial}{\partial \theta} \right] $, where $\alpha \equiv \frac{\hbar^2}{4 m_e^2 c^2} \frac{d}{d r} V(r)$, with $r_1$ the radius of the center line of torus, $r_0$ the radius of its cross-section, $\theta$ the poloidal angle. We shift our perspective to the circular helical structure, since what we are truly concerned with is the motion of electrons along the helical direction. Considering this property, $\mathcal{H}_{\mathrm{so}}$ can be rewritten as $\mathcal{H}_{\mathrm{so}} = \frac{\alpha}{2 \hbar}\left[\sigma_{t} \hat{p}_{s}+\hat{p}_{s} \sigma_{t}\right]$, where $\hat{p}_{s}$ is the momentum direction of electrons propagating along the CSH, $\sigma_t(\theta, \varphi)=\cos \theta \sin \beta \sigma_x+(\sin \varphi \cos \beta-\sin \theta \cos \varphi \sin \beta) \sigma_y-(\sin \theta \sin \varphi \sin \beta+\cos \varphi \cos \beta) \sigma_z$ with $\beta$ the space angel and $\varphi$ the toroidal angle. Similarly, by transforming into the second quantization representation, $\mathcal{H}_{\mathrm{so}}$ can be written as
\begin{eqnarray}
\label{soc Hamiltonian}
\mathcal{H} _{so}=\sum_{n=1}^{\mathcal{N}-1} i \lambda_{so} c_n^{\dagger}\left(\sigma_n^m+\sigma_{n+1}^{m+1}\right) c_{n+1}+\text { H.c.},
\end{eqnarray}
where $\lambda_{so} = -\frac{\alpha}{4 s}$ and $\quad \sigma_{n+1}^{m+1}=\sigma_{t}(m\Delta \theta,n \Delta \varphi)$, with $s$ the arc length, $m = n$ mod $M$, $\Delta \theta= 2\pi /M$, $\Delta \varphi= 2\pi /\mathcal{N}$, and $M$ the number of atoms per lap.

The third term $\mathcal{H}_{l}$ $(=\sum_{k, \beta}(\varepsilon_0 a_{\beta k}^{\dagger} a_{\beta k}+t_0 a_{\beta k+1}^{\dagger} a_{\beta k}+\text { H.c.}))$ represents the Hamiltonian for the electrons in the left and right leads. The fourth term $\mathcal{H}_{lc}$ (=$\sum_{\beta}(\gamma_{\beta}a_{\beta1}^{\dagger}c_{n_{\beta}}+\mathrm{H}.\mathrm{c}.))$ indicates the coupling between the CSH molecule and both leads. Note that this term plays a vital role in facilitating the charges' transferring in the system and in above equations, $\beta = L, R$, $n_L = 1$, $n_R = N$, $\epsilon_0$ denotes the on-site energy, $t_0$ is the hopping strength for leads and $a_{\beta k}^{\dagger}$ corresponds to the creation operations of the $k^{th}$ site in the left and right leads.


Additionally, the last two terms, $\mathcal{H}_p+\mathcal{H}_{p c}$, indicate respectively the Hamiltonian of the voltage probe used to simulate incoherent-inelastic phenomena and the coupling term between
the probe and the central CSH molecule. To gain the academic purpose of this study, we simulate each probe as an one-dimensional (1D) ordered semi-infinite chain. These two terms are specified by $\mathcal{H}_p=\sum_{n, l}(\varepsilon_0 d_{n l}^{\dagger} d_{n l}+t_0 d_{n l+1}^{\dagger} d_{n l}+\text { H.c.})$ and
$\mathcal{H}_{p c}=\sum_{n}(\gamma c_n^{\dagger} d_{n 1}+\text {H.c.})$, where the on-site energy $\epsilon_0$ and the hopping parameter $t_0$ are set in the same way as we performed in both leads. With the help of the coupling $\gamma$, the first site of every probe is linked to the $n$-th site of the central CSH molecule.

\subsection{B\"{u}ttiker's probe technique}

The Landa\"{u}er transport theory offers a straightforward and accurate explanation of quantum transport with phase coherence \cite{Datta}. Its straight forwardness makes it attractive to extend its application beyond the coherent threshold \cite{Buttiker1985,Buttiker1986}. We can incorporate environmental impacts like phase loss, dissipation and energy exchange into electron transport calculations with the aid of the LBP approach. In this work, we provide an overview of this technique, beginning with a specifical and broad explanation. To this end, we apply the voltage probe method as outlined in \cite{Kilgour2015}.

Our model Hamiltonian described in Eq.(1) follows the Landauer-B\"{u}ttiker formula for the electrical current exiting the $\nu$ contact, because it does not incorporate explicit many-body interactions,
\begin{gather}
I_{\nu}=\frac{e}{h}\sum_{\alpha \sigma}{\int_{-\infty}^{\infty}{d}}\epsilon \mathcal{T} _{\nu ,\alpha}^{\sigma}(\epsilon )\left[f_{\nu}(\epsilon )-f_{\alpha}(\epsilon)\right],\quad \nu =L,R,
\end{gather}
where the Fermi-Dirac distribution functions in the $\nu$ terminal are expressed using the inverse temperature $\beta_{\nu} = (k_B T_v)^{-1}$ and chemical potential $\mu$, and denoted by $f_{\nu}(\epsilon)=\left[e^{\beta_{\nu}\left(\epsilon-\mu_{\nu}\right)}+1\right]^{-1}$. The determination of the distribution functions $f_{\nu}(\epsilon)$ is based on the voltage probe requirements, which are explained below. Following the same logic as in Eq. (4), the Landauer-B\"{u}ttiker formula provides the magnitude of the electrical current flowing out of the $n$-th probe as below
\begin{gather}
I_n=\frac{e}{h} \sum_{\alpha \sigma} \int_{-\infty}^{\infty} d \epsilon \mathcal{T}_{n, \alpha}^{\sigma}(\epsilon)\left[f_n(\epsilon)-f_\alpha(\epsilon)\right].
\end{gather}
Note that the spin-dependent transmission coefficient in Eqs. (4) and (5) can be obtained from the molecular Green's function and the related hybridization matrices,
\begin{gather}
\mathcal{T}_{\alpha, \alpha^{\prime}}^{\sigma}(\epsilon)=\operatorname{Tr}\left[\Gamma_{\alpha^{\prime}}(\epsilon) G^r(\epsilon) \Gamma_\alpha(\epsilon) G^a(\epsilon)\right]^{\sigma},
\end{gather}
where the trace is performed over the $N$ electronic states of the CSH molecule. In the present wide-band limit, the matrix components of  $\Gamma_\alpha$ are constants by real values, without considering the energy dependence. Note that the retarded Green's function is given by
\begin{gather}
G^r(\epsilon)=\left[\epsilon I -\mathcal{H}_{c}-\mathcal{H}_{\mathrm{so}}+i \Gamma / 2\right]^{-1},
\end{gather}
with $G^a(\epsilon)=\left[G^r(\epsilon)\right]^{\dagger}$ and $\Gamma=\Gamma_L+\Gamma_R+\sum_{n=1}^\mathcal{N} \Gamma_n$. Within the pertinent configurations, the molecule is attached to each metallic lead by means of a solitary site, whereby the left (right) lead is connected to site "1" ("$\mathcal{N}$"). Hence, the hybridization matrices comprise one solitary non-zero value in the following manner,
\begin{gather}
\left[\Gamma_n\right]_{n, n}=\gamma_d,\
\left[\Gamma_L\right]_{1,1}=\gamma_L,\
\left[\Gamma_R\right]_{\mathcal{N}, \mathcal{N}}=\gamma_R,
\end{gather}
with the energy parameters $\gamma_{L,R}$ and $\gamma_d$, describing the coupling between the central molecule and the metal leads, and the coupling between the $n$-th probe with the helical molecule.

The voltage probes not only cause a minor broadening of molecular states, but also exhibit an incoherent scattering effect which extends beyond the coherent Landa\"{u}er model. the conductive electrons have the ability to transfer from the center molecular to the voltage probe due to the coupling between them. Although we ensure that no current leaks to either probe, i.e., $I_n=0$ for $n=1,2, \ldots, \mathcal{N}$ shown in Eq. (5), the electrons lose their phase coherence in the probes and reintegrate with the molecule at various energy levels dictated by the probe's energy spectrum.

We enforce the current condition of zero charge for every probe, which subsequently determines the local chemical potential $\mu_n$ of each probe. Explicitly, $I_n=0$ constitutes a set of nonlinear equations for $\mu_n$,
\begin{equation}
\begin{aligned}
\sum_{\alpha \sigma} \int_{-\infty}^{\infty}d \epsilon \mathcal{T}_{n, \alpha}^{\sigma}(\epsilon) f_n(\epsilon) = & \sum_{n^{\prime} \sigma} \int_{-\infty}^{\infty}d \epsilon \mathcal{T}_{n, n^{\prime}}^{\sigma}(\epsilon) f_{n^{\prime}}(\epsilon)  \\
& +\sum_{\nu \sigma} \int_{-\infty}^{\infty} d \epsilon\mathcal{T}_{n, \nu}^{\sigma}(\epsilon) f_{\nu}(\epsilon).
\end{aligned}
\end{equation}
The above formula comprises a group of $N$ nonlinear equations, making it challenging to obtain an accurate analytical solution. Thus we need deal with this issue with the help of a purely numerical approach, such as the Newton-Raphson method. But here, for convenience, we confine our calculations in the linear response regime. Assuming low voltage, we can expand the Fermi-Dirac functions using Taylor series up to the first non-trivial order,
\begin{gather}
f_\alpha\left(\epsilon, \mu_\alpha\right)=f_{e q}\left(\epsilon, \epsilon_F\right)-\frac{\partial f_{e q}\left(\epsilon, \epsilon_F\right)}{\partial \epsilon}\left(\mu_\alpha-\epsilon_F\right).
\end{gather}

Then, we explicitly indicate the dependency of the Fermi function on the
Fermi energy $\epsilon_F = \mu_{eq}$ and for convenience, we set $\epsilon_F = 0$ below. In the linear response regime, there is minimal energy transfer between the CSH molecule and the environment, causing the voltage probe to function like a dephasing probe. However, when far from equilibrium, the voltage probe inelastically scatters electrons among molecular orbitals, absorbing electrons and subsequently injecting them with a variety of energies, which are are determined by the distribution function within each probe. With this treatment, Eq. (9) can be reduced into a set of $\mathcal{N}$ linear equations as written below
\begin{equation}
\begin{aligned}
& \mu_n \sum_{\alpha \sigma} \int_{-\infty}^{\infty}\left(-\frac{\partial f_{e q}}{\partial \epsilon}\right) \mathcal{T}_{n, \alpha}^{\sigma}(\epsilon) d \epsilon \\
& -\sum_{n^{\prime}, \sigma}^N \mu_{n^{\prime}} \int_{-\infty}^{\infty}\left(-\frac{\partial f_{e q}}{\partial \epsilon}\right) \mathcal{T}_{n, n^{\prime}}^{\sigma}(\epsilon) d \epsilon \\
& =\int_{-\infty}^{\infty} d \epsilon\left(-\frac{\partial f_{e q}}{\partial \epsilon}\right)\left[\mathcal{T}_{n, L}^{\sigma}(\epsilon) \mu_L+\mathcal{T}_{n, R}^{\sigma}(\epsilon) \mu_R\right].
\end{aligned}
\end{equation}
The above set of $N$ equations can be reformulated as a matrix equation $\boldsymbol{M}\cdot \boldsymbol{\mu}=\mathbf{v}$, which can be solved easily by a matrix inversion. This solution will yield the chemical potentials of the probes located at each site. Here the matrix $\boldsymbol{M}$ can be given by the following formula.

\begin{widetext}
$\boldsymbol{M}=\begin{pmatrix}
\sum_{\alpha \neq 1, \sigma} \int_{-\infty}^{\infty} \mathcal{T}_{1 \alpha}^\sigma(\epsilon)\left(-\frac{\partial f_{\mathrm{cq}}(\epsilon)}{\partial \epsilon}\right) d \epsilon & -\sum_{\sigma}\int_{-\infty}^{\infty} \mathcal{T}_{12}^\sigma(\epsilon)\left(-\frac{\partial f_{\mathrm{cq}}(\epsilon}{\partial \epsilon}\right) d \epsilon & -\sum_{\sigma}\int_{-\infty}^{\infty} \mathcal{T}_{13}^\sigma(\epsilon)\left(-\frac{\partial f_{\mathrm{cq}}(\epsilon)}{\partial \epsilon}\right) d \epsilon & \ldots \\
-\sum_{\sigma}\int_{-\infty}^{\infty} \mathcal{T}_{21}^\sigma(\epsilon)\left(-\frac{\partial f_{\mathrm{cq}}(\epsilon)}{\partial \epsilon}\right) d \epsilon & \sum_{\alpha \neq 2, \sigma} \int_{-\infty}^{\infty} \mathcal{T}_{2 \alpha}^\sigma(\epsilon)\left(-\frac{\partial f_{\mathrm{cq}}(\epsilon)}{\partial \epsilon}\right) d \epsilon &-\sum_{\sigma} \int_{-\infty}^{\infty} \mathcal{T}_{23}^\sigma(\epsilon)\left(-\frac{\partial f_{\mathrm{cq}}(\epsilon)}{\partial \epsilon}\right) d \epsilon & \ldots \\
-\sum_{\sigma}\int_{-\infty}^{\infty} \mathcal{T}_{31}^\sigma(\epsilon)\left(-\frac{\partial f_{\mathrm{eq}}(\epsilon)}{\partial \epsilon}\right) d \epsilon &-\sum_{\sigma} \int_{-\infty}^{\infty} \mathcal{T}_{32}^\sigma(\epsilon)\left(-\frac{\partial f_{\mathrm{cq}}(\epsilon)}{\partial \epsilon}\right) d \epsilon & \sum_{\alpha \neq 3, \sigma} \int_{-\infty}^{\infty} \mathcal{T}_{3 \alpha}^\sigma(\epsilon)\left(-\frac{\partial f_{\mathrm{cq}}(\epsilon)}{\partial \epsilon}\right) d \epsilon & \ldots \\
\ldots & \ldots & \ldots & \ldots
\end{pmatrix}$.
\end{widetext}
Using further the chemical potentials of the probes, we linearize Eq. (4) to achieve the net charge current that flows through the CSH molecule in a linear response,
\begin{gather}
I_L=\frac{e}{h}\sum_{n \sigma}{\left[ \int_{-\infty}^{\infty}{\mathcal{T} _{L,n}^{\sigma}}(\epsilon )\left( -\frac{\partial f_{eq}}{\partial \epsilon} \right) d\epsilon \right]}\left( \mu _L-\mu _{n} \right).
\end{gather}

\begin{figure*}
\includegraphics[width=6.50in]{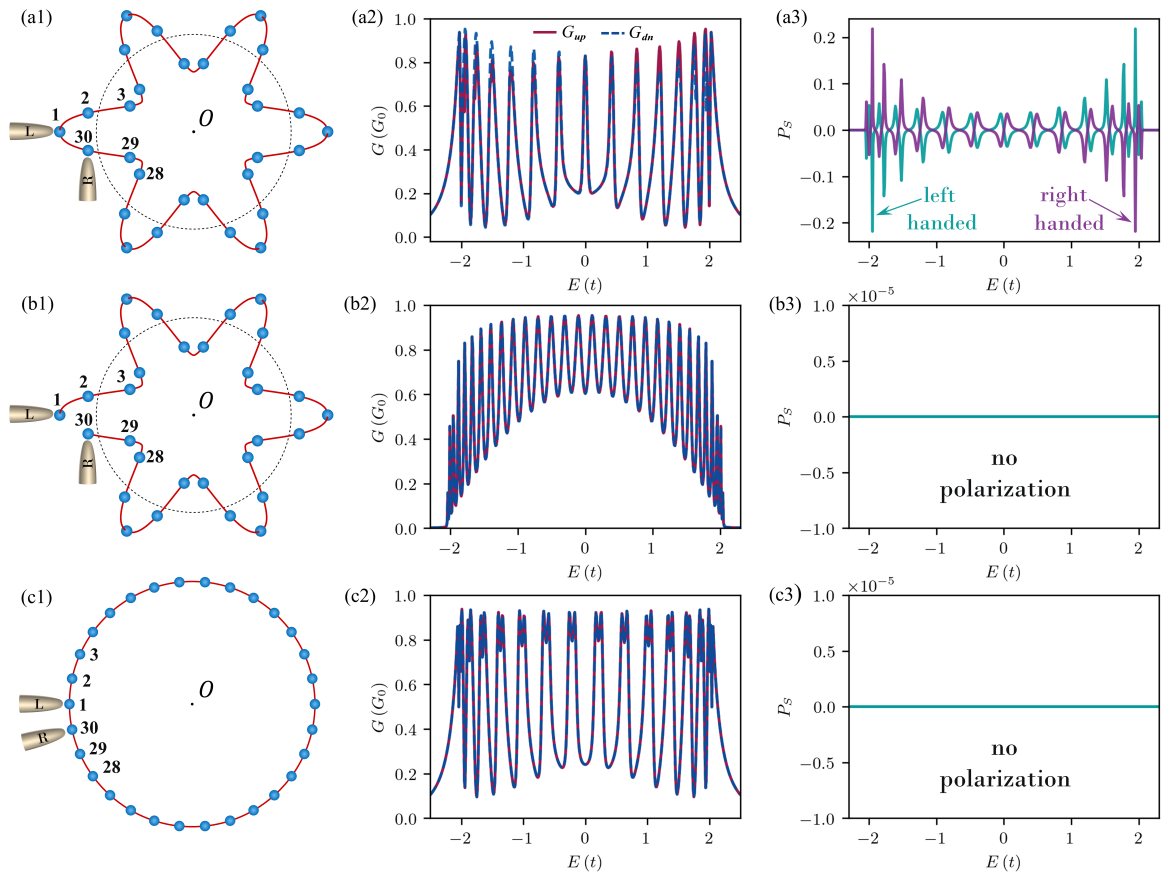}
\caption{(a1) Schematic top view of a connected circular single helical molecule with the left electrode connected to the first site and the right electrode connected to the $\mathcal{N}$-th site ($\mathcal{N} = 30$). (a2) Spin-up conductance $G_{up}$ (red color), spin-dn one $G_{dn}$ (blue color) and (a3) spin polarization $P_{S}$ (cyan color) as a functions of incoming electron energy $E$ for a circular left-handed molecules, where the purple line represents the $P_{S}$ of right-handed molecules. Schematic top view,  $G_{up(dn)}$ and  $P_{S}$ in the absence (b1)-(b3) of the connection circular structures and (c1)-(c3) of the helical symmetry. Here, $G_0 = e^2 / 2 \pi \hbar$ is the universal quantum electrical conductance. The parameters are $\mathcal{N} = 30$, $\lambda_{so}= 0.1t$, $\gamma_L = \gamma_R = t$, $\gamma_d = 0.005t$.}
\label{fig2}
\end{figure*}

The technique of using a voltage probe produces the electric conductance of the helical molecule junction within a linear response. $G=I_L / \Delta V \text { with } \Delta V=\left(\mu_R-\mu_L\right) / e$ being the applied bias voltage. When examining Eq.(12) under conditions of low temperature and voltage bias, the distinction between Fermi functions decreases to a Dirac delta function, and then the conductance can be expressed as

\begin{gather}
\label{conductance}
G_{\sigma}=\frac{e^2}{h}\left( \mathcal{T} _{L,R}^{\sigma}\left( \epsilon _F \right) +\sum_{n \sigma}{\mathcal{T} _{L,n}^{\sigma}}\left( \epsilon _F \right) \left( \mu _L-\mu _n \right) /\Delta \mu \right).
\end{gather}

\section{RESULTS AND DISCUSSION}

To ensure the accuracy of numerical calculations, the structural parameters are deemed uniformly throughout every coil of the central CSH molecules. For the CSH molecule considered here, the atoms in per lap are adopted as $M =5$ and the remaining geometric structural parameters are provided in the Supplementary Material (SM) \cite{SpM}. The on-site energy is set as $\varepsilon_n = 0$ without loss of generality, the $NN$ hopping integral $t$ of electrons between two sites is taken as the energy unit, and the SOC is estimated to be $\lambda_{so}= 0.1t$. Unless otherwise stated, for the real electrodes and virtual leads, the parameters $\gamma_L = \gamma_R = t$ and $\gamma_d = 0.005t$ are adopted.

Initially, we explore the occurrence of CISS effect in the CSH molecule with the length $\mathcal{N} = 30$, in which the source and drain electrodes are contacted with the first and $\mathcal{N}$-th site position as drawn in Fig.~\ref {fig2}(a1). The calculated spin-up and spin-down conductances with respect to the incoming electron energy $E$ in the left-handed molecule are illustrated in Fig.~\ref{fig2}(a2). It is observed that the conductance curve exhibits oscillating behaviors. Our model involves a circular geometry connected with two leads, in which the wave functions of incoming electrons traverse through both arms of the circular structure. Subsequently, they converge at the junction in which the lead 2 is attached to the central CSH molecule, resulting in constructive quantum interference (CQI) or destructive quantum interference (DQI), and the former gives rise to oscillations in the conductance. The number of resonant peaks in the oscillation is equivalent to that of energy levels in the isolated CSH molecule. Notably, the incoming electrons are entirely unpolarized due to the non-magnetic leads applied. More importantly, the unequal conductance between spin-up and spin-down states associated with energy indicates the occurrence of spin polarization (see the cyan color in Fig.~\ref{fig2}(a3)). Interestingly, when we adjust the chirality of the CSH molecule by changing the twist and space angles from $\theta$ to -$\theta$ and $\beta$ to $\pi$-$\beta$, a right-handed molecule is achieved here. It is clear that the spin polarization is changed to an opposite direction due to the altering handedness of the CSH molecule, as described by the purple line in Fig.~\ref{fig2}(a3).

\begin{figure*}
\includegraphics[width=2.05\columnwidth]{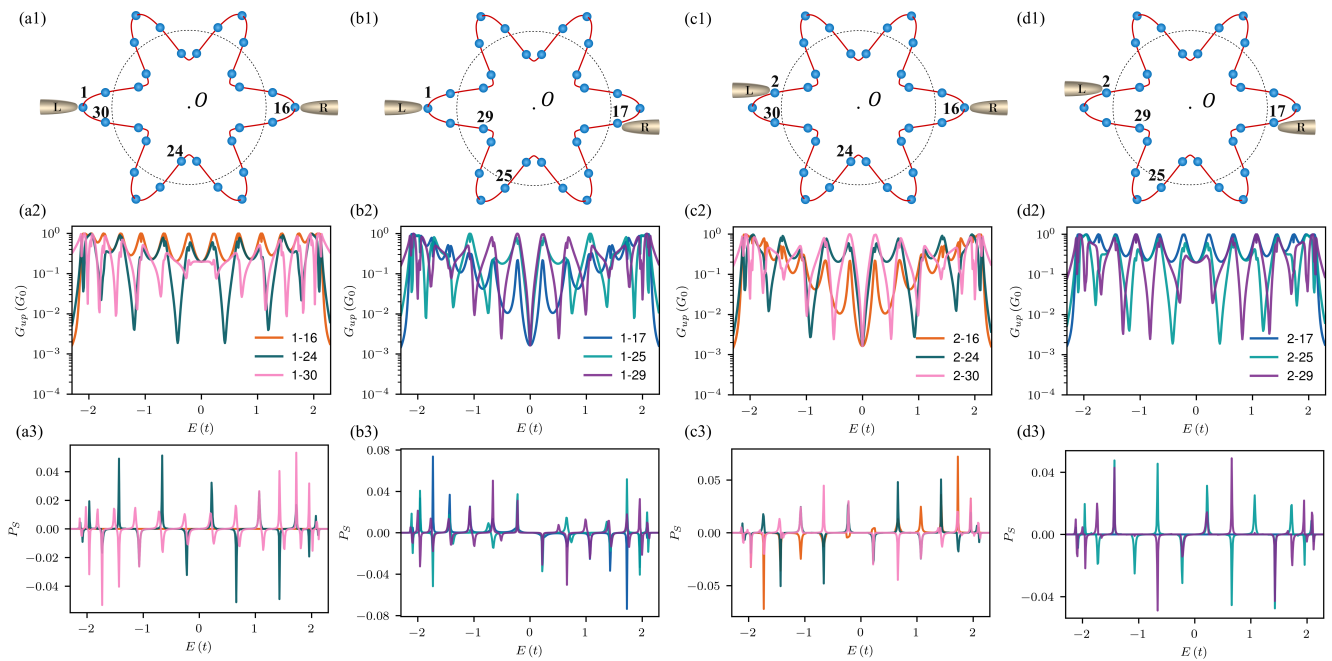}
\caption{The device schematic and the related spin-dependent transport properties of a circular single helix molecule with the length $N = 30$, in which the left electrode is connected with the first lattice site, while the right electrode is connected with even (a1-a3) and odd (b1-b3) lattice sites, and the left electrode is connected to the second lattice site, while the right electrode is connected to even (c1-c3) and odd (d1-d3) lattice sites, other parameters are $\lambda_{so} = 0.12t$ and $\gamma_d = 0.001t$.}

\label{fig3}
\end{figure*}

In order to determine the effective factors associated with spin polarization phenomena in CSH molecules, in Fig.~\ref{fig2}(b1)-(b3) and (c1)-(c3), we present the top view of CSH-based device and the spin-dependent conductances under the conditions in which the connection of CSH geometry and helical symmetry are absent, respectively. It is evident that the spin polarization disappears when these two factors are eliminated. When we consider that electrons no longer undergo hopping between the 1-st and 30-th site, the resulting circular structure is topologically equivalent to a helical chain. In this case, we can obtain a spin-independent Hamiltonian through unitary transformation to ensure strictly zero spin polarization (see Fig.~\ref{fig2}(b3)), due to the uniqueness of the electron transport path. on the contrary, the existence of counterclockwise and clockwise transport pathways in CSH molecules, similar to the base pair transport pathways in double-stranded DNA and the long-range hopping transport pathways in proteins \cite{GuoSun2012, GuoSun2014}, leads to the spin polarization phenomenon observed in CSH molecules (see Fig.~\ref{fig2}(a3)). Note that in single-stranded DNA, due to the specific geometric structure, its long-range hopping is much smaller compared to the NN hopping, which makes it unable to generate an obvious spin polarization \cite{GuoSun2014,Gohler2011}. However, our model suggests that when the single-stranded DNA is connected end to end to construct a circular one, the quantum coherence tends to result in spin polarization without the influence of long-range hopping. Further, when helical symmetry is absent, it is evident that $P_{S}$ maintains zero, regardless of the values of other parameters, as illustrated in Figs.~\ref{fig2}(c2) and (c3), in which $G_{up(dn)}$ and $P_{S}$ are depicted and the relations $\Delta \theta=0$ and $\beta_m=\pi/2$ are adopted. Nevertheless, an interesting case is that the spin-polarization direction (i.e., the \emph{z}-axis plotted in Fig.~\ref{fig1}(a)) in our device model is chosen to be parallel with the molecular plane, which results in a completely different result in the absence of chirality in comparison with a mesoscopic rings with Rashba spin-orbit coupling \cite{mesoscopic_rings2003, mesoscopic_rings2002}, and the latter exhibits an obvious spin polarization. Finally, previous literatures reported that the CISS can be observed in a paramagnetic phase \cite{InuiPRL2020, Shishido2021}, indicating that spin polarization emerges in a specific state possessing the time-reversal symmetry. However, it is well-established that a spin current proportional to the electric field does not appear in the single-channel systems which possess time-reversal symmetry \cite{Bardarson2008}. Therefore, when the voltage probe is removed from the CSH molecule, the spin polarization disappears from the system due to the absence of non-unitary effects inside the chiral molecule, see SM, Fig. S1 for more details \cite{SpM}.

\begin{table*}[!t]
\caption{The spin-up orbital energies contributions to $\Gamma_{r,s}(E_F)$ for  circular single helical molecules whose length is $N = 30$. Here, $\Gamma_{r,s}(E_F)=\sum_i \varGamma _{r,s}^{i}\left( E_F \right)$
, with $\varGamma _{r,s}^{i}\left( E_F \right)$ the contributions to the i-th molecules orbital and $\Gamma_{r,s}(E_F)$ the contribution of all molecular orbitals where $r$ and $s$ represent indices for the positions where the left and right electrodes couple to the molecule, respectively.}\label{molecular orbitals}
\begin{tabular}{p{1.4cm}|p{1.2cm}|p{1.2cm}|p{1.2cm}|p{1.2cm}|p{1.2cm}|p{1.2cm}|p{1.2cm}|p{1.2cm}|p{1.2cm}|p{1.2cm}|p{1.2cm}|p{1.2cm}}
\hline
\hline
\makecell[c]{$\varGamma _{r,s}^{i}\left( E_F \right)$} &\makecell[c]{$\varGamma _{1,16}^{i}$} &\makecell[c]{$\varGamma _{1,17}^{i}$} &\makecell[c]{$\varGamma _{1,24}^{i}$} &\makecell[c]{$\varGamma _{1,25}^{i}$}  &\makecell[c]{$\varGamma _{1,29}^{i}$} &\makecell[c]{$\varGamma _{1,30}^{i}$}  &\makecell[c]{$\varGamma _{2,16}^{i}$} &\makecell[c]{$\varGamma _{2,17}^{i}$}&\makecell[c]{$\varGamma _{2,24}^{i}$}&\makecell[c]{$\varGamma _{2,25}^{i}$}&\makecell[c]{$\varGamma _{2,29}^{i}$}&\makecell[c]{$\varGamma _{2,30}^{i}$}\\
\hline
\makecell[c]{LUMO} & \makecell[c]{-0.1702}  &\makecell[c]{0.0128} &\makecell[c]{0.0734}   &\makecell[c]{0.0944} & \makecell[c]{0.1434} & \makecell[c]{0.0583} & \makecell[c]{0.0552}& \makecell[c]{-0.0054}& \makecell[c]{-0.1044}& \makecell[c]{-0.1186}& \makecell[c]{-0.1295}& \makecell[c]{-0.0792}\\
\hline
\makecell[c]{HOMO} & \makecell[c]{-0.1702}  &\makecell[c]{-0.0128} &\makecell[c]{0.0734}   &\makecell[c]{-0.0944} & \makecell[c]{-0.1434} & \makecell[c]{0.0583} & \makecell[c]{-0.0552}& \makecell[c]{-0.0054}& \makecell[c]{0.1044}& \makecell[c]{-0.1186}& \makecell[c]{-0.1295}& \makecell[c]{0.0792}\\
\hline
\hline
\makecell[c]{$\varGamma _{r,s}\left( E_F \right)$} & \makecell[c]{-0.4189}  &\makecell[c]{0} &\makecell[c]{-0.0118}   &\makecell[c]{0} & \makecell[c]{0} & \makecell[c]{0.0895} & \makecell[c]{0}& \makecell[c]{-0.0266}& \makecell[c]{0}& \makecell[c]{-0.2096}& \makecell[c]{-0.2302}& \makecell[c]{0}\\
\hline
\hline
\end{tabular}
\end{table*}

To elucidate the CISS effect and its accompanying spin-dependent DQI effect under different molecular geometries and with various contacting positions in CSH molecules, we first investigate the influences of molecular lengths and the positions of the left and right electrodes on the spin-polarized conductance. The top-down schematic diagram of devices, the related spin-up conductance and spin polarization of the CSH molecule with $N=30$ are illustrated in Fig.~\ref{fig3}, in which the left and right electrodes are connected with different lattice positions. At the first glance, a clear dip in the spin-dependent conductance can be observed at the Fermi level when the two connection sites are situated on the same sublattice, specifically the odd-odd, even-even sublattices (see Fig.~\ref{fig3}(b2) and (c2), respectively). This dip is located at the midpoint of the HOMO-LUMO gap and provides the direct evidence for the existence of DQI. In contrast, when the connection sites are placed on different sublattices, specifically the odd-even, even-odd (see Figs.~\ref{fig3}(a2) and (d2)), the conductance remains relatively smooth around the Fermi level, suggesting the presence of CQI. This is in a good agreement with the result appearing in other bipartite carbon systems \cite{ulvcakar2019resilience, Sangtarash,Geng,Sangtarash4469}. Similarly, for other systems with the length of $\mathcal{N}$, where $\mathcal{N} \bmod 2=0 \wedge \mathcal{N}\bmod M=0$, their conductance curves also satisfy the above conclusion (see SM \cite{SpM}). In addition, if we consider the case in which $\mathcal{N} \bmod 2 =1 \wedge \mathcal{N} \bmod M=0$, the energy-dependent conductance displays a notable distinction from the above scenarios. Specifically, regardless of whether the two connecting sites are located within the same or different sublattices, the distinct sharp valley near the Fermi level observed in those cases in which $\mathcal{N} = 30$ is absent, as provided in SM \cite{SpM}.

On the other hand, the component of spin polarization along the axis of spin quantization denotes the normalized difference of the spin-up and spin-down electrons transport through the right lead. In particular, the spin-up electrons and the spin-down ones exhibit oppositive scattering behaviors while moving through the central molecule under the influence of SOC, as demonstrated in spin Hall effect, in which the accumulation of different spins occurs at the opposite edges. Therefore, by adopting the line connecting the left electrode and the central point $O$ of the molecule as the axis of symmetry, the opposite spin polarizations are obtained when the right electrode is connected to symmetric positions in turn (see  Fig. S3 in SM for more details \cite{SpM}). Moreover, this feature also indicates that if the positions connected with the right electrode are collinear with the positions of the left electrode and the point $O$ for the even $\mathcal{N}$. For example, when $\mathcal{N}=30$, the left electrode is connected with the first and second sites, and the right electrode with the 16th and 17th sites, resulting in a zero spin polarization, as shown in Figs.~\ref{fig3}(a3) and (d3). In addition, due to the quantum interference from the travelling electrons in different pathes, the oscillating behavior appears in the spin polarization versus energy, then we may write the number of zero points (where the spin polarization is equal to 0) corresponding to spin-polarized function as $2\left|\mathcal{N}_R-\mathcal{N}_L\right|-\mathcal{N}-1$.
Here, $\left|\mathcal{N}_R-\mathcal{N}_L\right|$ represents the site difference between the left and right electrodes, and satisfies $\left|\mathcal{N}_R-\mathcal{N}_L\right| \geqslant  \frac{\mathcal{N}+2}{2}$  for even $\mathcal{N}$, and $\left|\mathcal{N}_R-\mathcal{N}_L\right| \geqslant  \frac{\mathcal{N}+1}{2}$ for odd $\mathcal{N}$.


Further, the above-mentioned particular transport behaviors can be understood by examing the related Green's function contributed from distinct molecular orbitals, as evidenced in previous literatures \cite{Taniguchi11426, Yoshizawa1612,Stadler2017}. As for the quantum interference phenomena occurring in the vicinity of Fermi level, the HOMO and LUMO molecular orbitals work as the primary contributors to the related Green's function. The detailed formulation of Green's function can be expressed below,

\begin{gather}
\label{Green_HOMO_LUMO}
G_{\mathrm{RL}}(E_{\mathrm{F}}) \approx \frac{C_{\mathrm{R}, \mathrm{HOMO}} C_{\mathrm{L}, \mathrm{HOMO}}^*}{E_{\mathrm{F}}-\varepsilon_{\mathrm{HOMO}}}+\frac{C_{\mathrm{R}, \mathrm{LUMO}} C_{\mathrm{L}, \mathrm{LUMO}}^*}{E_{\mathrm{F}}-\varepsilon_{\mathrm{LUMO}}},
\end{gather}
where the molecular-orbital expansion coefficient at the atom linked to the left electrode is denoted as $C_{\mathrm{L}, \mathrm{HOMO}}$ and $C_{\mathrm{L}, \mathrm{LUMO}}$ for the HOMO and LUMO, respectively. The remaining expressions are similarly allocated. $\varepsilon_{\mathrm{HOMO}}$ and $\varepsilon_{\mathrm{LUMO}}$ are the molecular orbital energies of the
HOMO and LUMO. In the event that these two terms exhibit contradictory signs, Green's function will exhibit a decrease, ultimately resulting in DQI. In contrast, if them are congruent, a CQI will be generated. The Fermi level $E_{\mathrm{F}}$ localized between HOMO and LUMO, produces opposite denominators in Eq.~(\ref{Green_HOMO_LUMO}). The quantum interference, therefore, is dictated by the signs of the numerators in both expressions. When the numerators display identical signs, the Green's function will decrease, resulting in the occurrence of DQI and conversely, the occurrence of CQI is anticipated.

\begin{figure}
\includegraphics[width=\columnwidth]{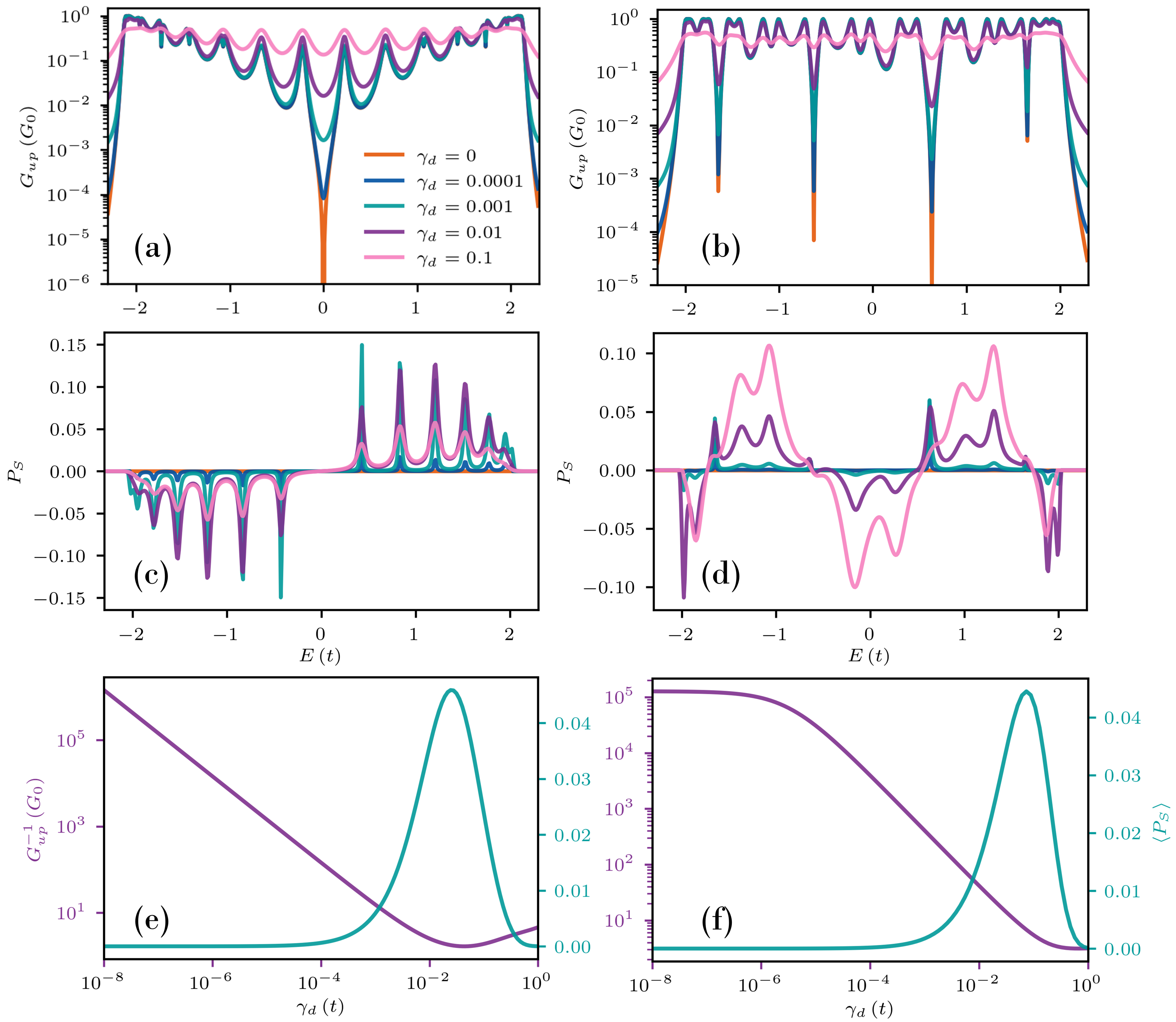}
\caption{The influence of dephasing strength $\gamma_d$ on the CISS and DQI effect in the CSH molecules while connecting the left lead to the 1st site and the right lead to the 17th site, which has a length of 30 ((a), (c), and (e)). Additionally, we connect the right lead to the 16th site, which has a length of 25 ((b), (d), and (f)). Here, (a) and (b) demonstrate the variation of the  spin-up conductance with different dephasing $\gamma_d$ in two particular cases. (c) and (d) demonstrate the variation of spin polarization at distinct dephasing strength $\gamma_d$ in two situations. (e) and (f) reveal the curves of the reciprocal of spin-up conductance $G_{up}^{-1}(E_F)$ and the average spin polarization $\left< P_S \right> $  as they are changed with $\gamma_d$ in two different scenarios.}
\label{fig4}
\end{figure}

However, this frontier molecular orbital approximation yields accurate outcomes only when the Coulson-Rushbrooke pairing theorem \cite{Coulson193, Gutman1986} can be utilized for forecasting DQI. It will become apparent that this one is universally applicable in situations where both the R-th and L-th atoms belong to either starred set or unstarred one. For all other instances, it is necessary to take into account all molecular orbitals present in the system. If an alternative hydrocarbon is exposed to atomic locations that belong to distinct subsets, where one subset is starred and the other is unstarred based on the CR framework, then even though the HOMO and LUMO only interfere constructively, the appendages linked to MOs that occupy lower energy levels and unoccupied MOs with higher energy levels may still annul the tails of the frontier orbitals at $E_{F}=0$, potentially resulting in the occurrence of DQI. An instance of the limitation of the frontier orbital rule is evident in the linkage between $C_2$ and $C_3$ atoms (hard-zero QI case) within the molecule of butadiene. Examining solely the coefficients of HOMO and LUMO orbitals, it is unreasonable to anticipate any cancellation effects between them. However, we should not disregard the inputs from the adjacent HOMO-1 and LUMO+1 orbitals, while observe that the HOMO and HOMO-1 undergo cancellation, as do the LUMO and LUMO+1. To put it differently, by taking into account the contributions from the neighboring orbitals, we can identify the occurrence of cancellations among the molecular orbitals.

It is worth highlighting that Stadler \emph{et al.} conducted a detailed investigation of quantum interference using Larsson's formula and emphasized that all molecular orbitals should be considered in order to accurately determine the QI from an MO standpoint \cite{Stadler2017}. Therefore, in Table \ref{molecular orbitals}, we show the contributions from all molecular orbitals for the selected structural parameters with $\mathcal{N} = 30$, $\lambda_{so} = 0.12t$ and $\gamma_d = 0.001t$, which we obtained by making use of Larsson's formula \cite{Larsson1981},
\begin{equation}
\Gamma(E)=\sum_i \frac{\alpha_i \beta_i}{E-\varepsilon_i},
\end{equation}
where $\varepsilon_i$ represents the eigenenergy associated with each MO, while $\alpha_i$ and $\beta_i$ denote the respective couplings of the MO with the left and right electrodes. Larsson's formula was initially proposed to define the transfer integral in the context of Marcus theory, which describes electron hopping \cite{Larsson1981, Ratner1990, Kastlunger2014}. However, it has been demonstrated very recently that the formula can be applied to estimate $T(E)$ as $T(E) \sim \Gamma^2(E)$ for coherent tunneling \cite{Stadler125401, Sautet4910}. Table \ref{molecular orbitals} shows the magnitude of the component values $\varGamma _{r,s}^{i}\left( E_F \right)$ corresponding to different MO and their total sum $\Gamma _{r,s}\left( E_F \right)$. Here, $r$ and $s$ respectively represent the indices of the positions at which the left and right electrodes are connected with the central molecule. There is a compelling evidence indicating that the positions of the left and right electrodes' positions corresponding to $\Gamma _{r,s}\left( E_F \right)$ equals zero and finite values, respectively. Table \ref{molecular orbitals} coincides with the main results associated with the conductance features observed in Fig.~\ref{fig3}. It is worth noting that the discussion above is based on the weak interaction regime, in which the full Green's function is approximated as a bare Green's function. So the numerical results obtained are accurate enough for us to qualitatively describe the behavior of spin-dependent conductance at the Fermi level by considering the coupling between the central molecule and the environment. Note that this physical mechanism is also applicable to understand the particular spin-dependent transport behaviors in the CSH molecules with the length $\mathcal{N}$ = 20 and 25.

\begin{figure}
\includegraphics[width=\columnwidth]{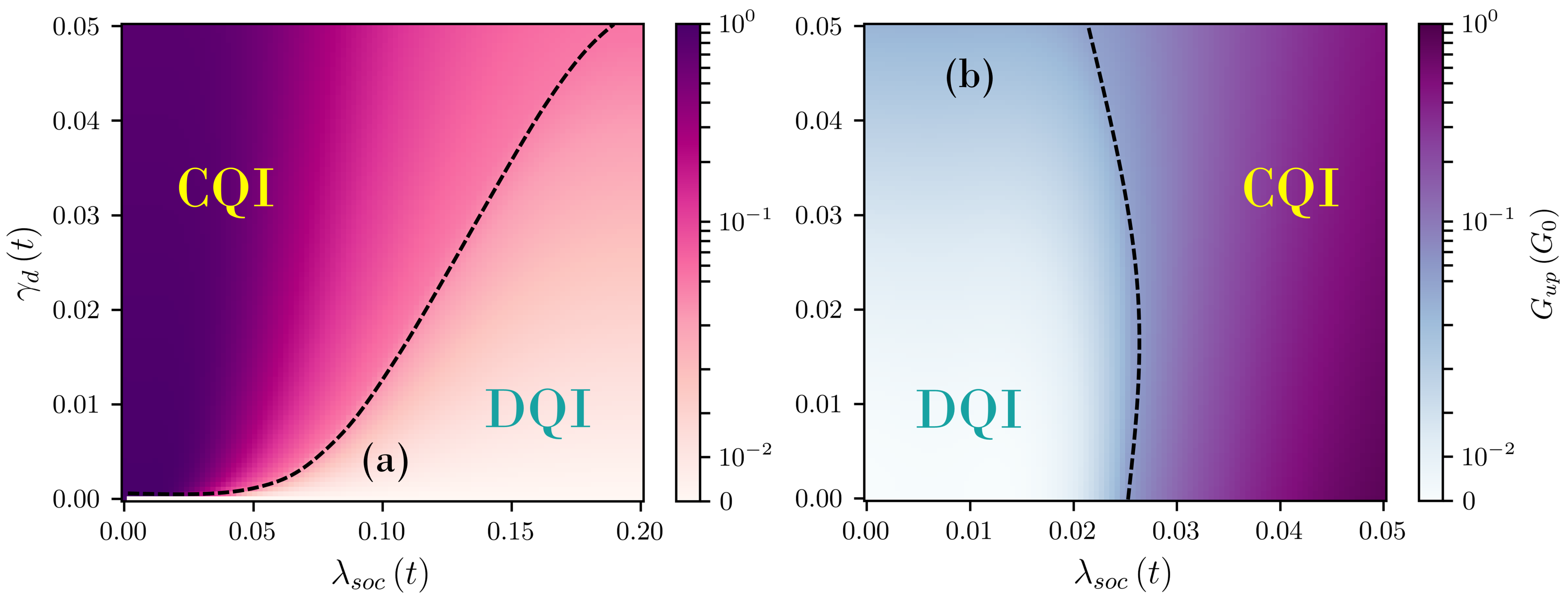}
\caption{The right electrode is connected to the 11th (a) and 12th (b) site, respectively, while the left electrode is connected to the first site. The spin-up conductance at the Fermi level $G_{up}(E_F)$ of a $\mathcal{N} = 20$ left-handed circular helical molecule is studied as a function of dephasing $\gamma_d$ and SOC $\lambda_{soc}$. }
\label{fig5}
\end{figure}

In what follows, we tend to study the impact of $\gamma_d$ on the CISS and DQI effect in the CSH molecules when the left electrode is connected to the 1st site and the right electrode is respectively connected to the 17th site with a length of 30 (see Figure (a), (c), and (e)), and connected to the 16th site with a length of 25 (see (b), (d), and (f)).
Figs.~\ref{fig4}(a) and (b) demonstrate the spin-dependent conductance of CSH molecules with different odd and even lengths. It is noteworthy noted that although the energies, where the quantum interference occurs, are different for these two cases, their anti-resonances are both enhanced as we weaken the strength of dephasing.
Moreover, Figs.~\ref{fig4}(c) and (d) demonstrate the spin-polarized transport behaviors corresponding to the occurrence of spin-dependent DQI effect in the CSH molecules. It can be seen that the intensity of spin polarization increases as the number of sites in the molecule increases, and the regularity of the number of spin polarization zero points corresponds to our description above. To study DQI and CISS comprehensively, we are now focusing on the reciprocal of the spin-up conductance $G_{up}^{-1}(E_F)$ and the average spin polarization $\left< P_S \right>$, where $\left. \langle P_s \right. \rangle \equiv \left( \left. \langle G_{\uparrow} \right. \rangle -\left. \langle G_{\downarrow} \right. \rangle \right) /\left( \left. \langle G_{\uparrow} \right. \rangle +\left. \langle G_{\downarrow} \right. \rangle \right) $ with $\left< G_{s} \right>$ averaged over the LUMO band for Fig.~\ref{fig4}(c), while over the energy range [-1.72, -0.63] for Fig.~\ref{fig4}(d). Fig.~\ref{fig4}(e) and (f) show that as the parameter $\gamma_d$ increases, $G_{up}^{-1}(E_F)$ decreases linearly overall and $\left< P_S \right>$ first increases and then decreases overall with increasing $\gamma_d$, regardless of whether $\mathcal{N}$ is odd or even. In addition, when $\mathcal{N}$ is even, $G_{up}^{-1}$ is larger than that when $\mathcal{N}$ is odd at very small $\gamma_d$ values, and the width of the $\left< P_S \right>$ curve when $\mathcal{N}$ is even is slightly larger than that when $\mathcal{N}$ is odd. As $\gamma_d$ gradually increases to around 1$t$, the overall conductance decreases and $G_{up}^{-1}$ rises due to the excessive coupling strength of the environment.

Finally, we focus our discussion on how the changes in the SOC $\lambda_{soc}$ and the dephasing strength $\gamma_d$ affect the DQI effect. Figs.~\ref{fig5}(a) and (b) demonstrate the spin-up conductance of a CSH molecule with the chain length of 20, with the left electrode connected with the 1st site, and the right electrode with the 11th and 12th site, respectively. For odd connecting positions, when the $\lambda_{soc}$ is small, even if $\gamma_d$ is very small but not equal to zero, it will cause the system to complete the phase transition from DQI to CQI. As the $\lambda_{soc}$ gradually increases, we also see the same phase transition with the increase of $\gamma_d$. However, for an even one, when the strength of $\gamma_d$ is within [0, 0.05], the phase transition of the system mainly depends on $\lambda_{soc}$ rather than $\gamma_d$. When SOC is small, the system is mainly localized in the DQI phase, while as SOC increases, the system undergoes a phase transition from DQI to CQI. Furthermore, considering the influence of SOC, when the length $\mathcal{N}$ of the CSH moelcule satisfies $\mathcal{N} \bmod 4 = 0 \wedge \mathcal{N} \bmod 10 = 0$, we can observe the phase transition between DQI and CQI at different odd and even electrode connection positions, as discussed above. However, when the length \emph{N} satisfies the relation $\mathcal{N} \bmod 4 = 1 \wedge \mathcal{N} \bmod 10 = 0$, no obvious phase transition can be observed. Interestingly, we find that for the CSH moelcule without any chirality and with other parameters unchanged, this phase transition also has no obvious manifestation, as provided in SM for more details \cite{SpM}.

\section{CONCLUSION}

In conclusion, we proposes a model Hamiltonian including the chirality-induced spin-orbit coupling and the environment-induced dephasing to investigate the effect of CISS and DQI on circular left-handed $\pi$-helix-like proteins connected to non-magnetic electrodes. Our theoretical study shows that the spin polarization effect is only observed when the central molecule possesses chirality, is in a connected state, and experiences environment-induced decoherence. The number of spin-polarized zero points at the Fermi level gradually increases depending on the number of lattice sites between the electrodes. The DQI effect weakens as the dephasing parameter increases, while the CISS effect first strengthens and then weakens. The phase transition between DQI and CQI can be observed only when some required conditions are satisfied.

\section{ACKNOWLEDGMENTS}

This work is supported by the National Natural Science Foundation of China with grant No. 11774104, 11504117, 11274128 and U20A2077, and partially by the National Key R\&D Program of China (2021YFC2202300).

\end{document}